\documentclass[twocolumn,showpacs,preprintnumbers,amsmath,amssymb]{revtex4-1}

\usepackage{xspace}
\usepackage{epsfig}
\usepackage{amsfonts}
\usepackage{graphicx}
\usepackage{url}
\usepackage{color}

\begin{document}

\preprint{AIP/123-QED}

\title{Polarization-induced Zener Tunnel Diodes in GaN/InGaN/GaN Heterojunctions} 


\author{Xiaodong Yan$^1$}
\author{Wenjun Li$^1$}
\author{S.M. Islam$^1$}
\author{Kasra Pourang$^1$}
\author{Huili (Grace) Xing$^{1,2}$}
\author{Patrick Fay$^{1}$}
\author{Debdeep Jena$^{1,2}$}
\email[]{djena@cornell.edu}
\affiliation{$^1$Department of Electrical Engineering, University of Notre Dame, Indiana 46556, USA.}
\affiliation{$^2$Departments of ECE and MSE, Cornell University, Ithaca, NY 14853, USA.}
\date{\today}

\begin{abstract}
By the insertion of thin In$_x$Ga$_{1-x}$N layers into Nitrogen-polar GaN p-n junctions, polarization-induced Zener tunnel junctions are studied.  The reverse-bias interband Zener tunneling current is found to be weakly temperature dependent, as opposed to the strongly temperature-dependent forward bias current.  This indicates tunneling as the primary reverse-bias current transport mechanism.  The Indium composition in the InGaN layer is systematically varied to demonstrate the increase in the interband tunneling current.  Comparing the experimentally measured tunneling currents to a model helps identify the specific challenges in potentially taking such junctions towards nitride-based polarization-induced tunneling field-effect transistors. 
\end{abstract}
\maketitle

Interband tunneling of electrons from filled states in the valence band to empty states in the conduction band occurs at high electric fields in uniformly doped semiconductors.  This was Zener's original proposal as the mechanism underlying dielectric breakdown \textcolor{red}{\cite{zenerPRCA34}}.  Esaki soon realized that the built-in electric field in the depletion region of a p-n junction can significantly boost tunneling - he used this principle to demonstrate NDR in Ge p-n junctions, as well as backward diodes \textcolor{red}{\cite{esakiPRB58, szeNgBook}}.  

Interband tunneling current in a reverse-biased p-n homojunction depends sensitively on the doping densities and the energy bandgap.  The doping densities $N_D$ and $N_A$ on the n-side and p-side control the depletion region thickness $W \sim \sqrt{\frac{2 \epsilon_s}{q} \frac{N_A + N_D}{N_A N_D} V_{bi}}$, which is roughly equal to the distance electrons need to tunnel at small negative biases.  Here $\epsilon_s$ is the dielectric constant of the semiconductor, $q$ is the electron charge, and $V_{bi} = \frac{kT}{q} \ln [ \frac{ N_A N_D }{n_{i}^{2}} ]$ is the built-in potential of the p-n junction, with $n_i$ the intrinsic carrier concentration generated by thermal interband excitations.  The built-in potential is close to the bandgap $E_g$, and the tunneling barrier height is roughly the energy bandgap. 

\begin{figure}[htbp]
	\centering
		\includegraphics[scale=0.6]{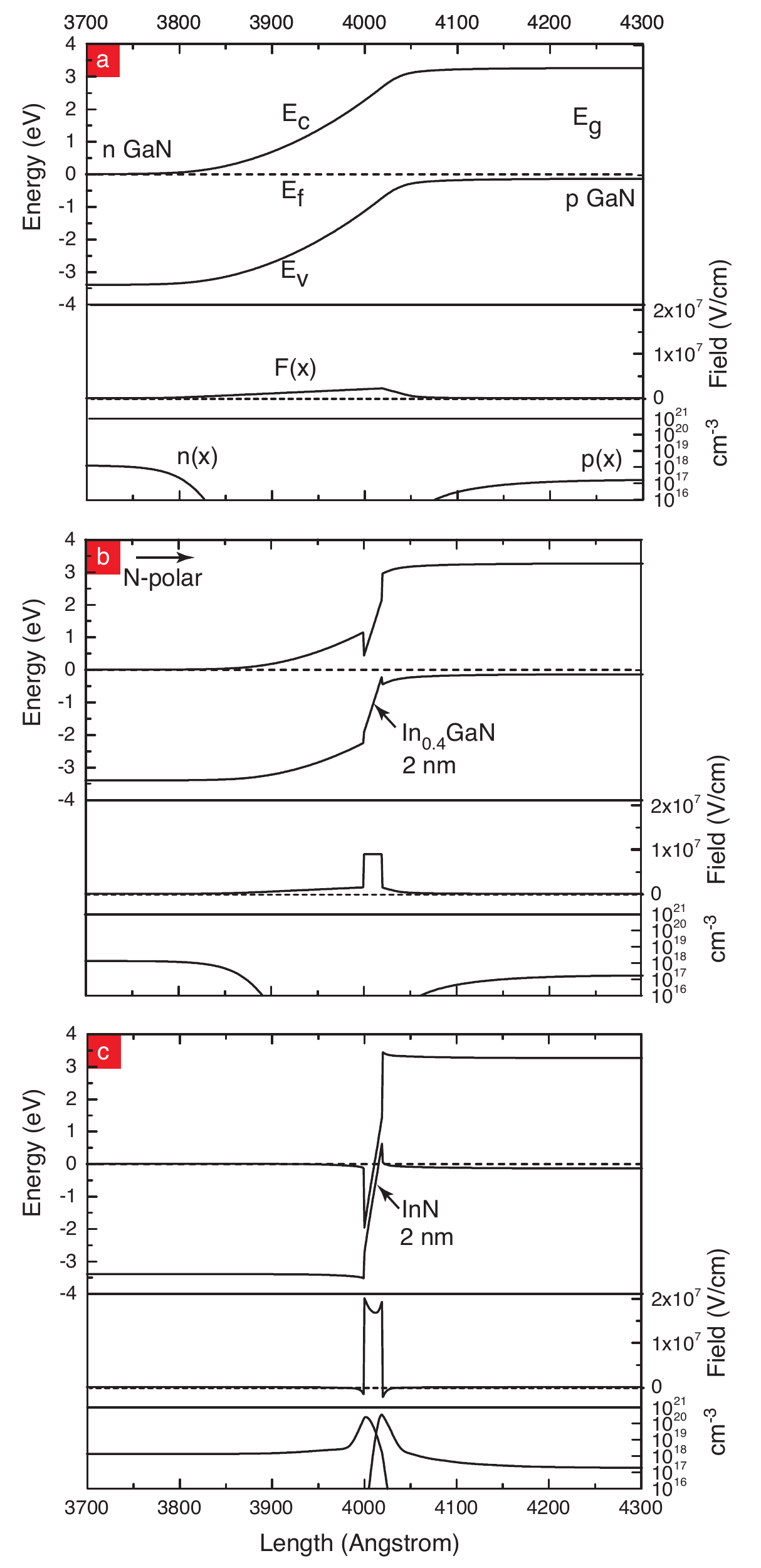}
	\caption{Calculated energy-band diagrams from self-consistent solutions of the Schrodinger and Poisson equations for (a) a GaN p-n junction, (b) with an InGaN polarization dipole, and (c) with an InN polarization dipole. The electric field profiles $F(x)$ and the mobile electron and hole distributions $n(x)$ and $p(x)$ are shown.  As the Indium content in the InGaN layer inceases, the effective band offset is tuned from Type-II (staggered) in (b) to Type-III (broken) in (c).  In going from (b) to (c), the junction is fundamentally changed from that of depletion to that of accumulation.}
	\label{fig1}
\end{figure}

The above standard p-n homojunction relations hold for shallow donor and acceptor levels.  Because of the deep-acceptor level of Mg p-type dopants in GaN, the above relations are modified significantly.  Figure \ref{fig1}(a) shows the calculated energy-band diagram, the electric field profile $F(x)$ and mobile carrier concentrations $n(x)$ and $p(x)$ for doping densities $N_D=5 \times 10^{18}$/cm$^3$ and $N_A=3 \times 10^{19}$/cm$^3$ at 300K.  The thermally ionized mobile hole concentration in the p-GaN far from the depletion region is only $p_p \sim 2 \times 10^{17}$/cm$^3$ because of a $E_A - E_V = 170 $ meV acceptor ionization energy, leading to a built-in voltage $V_{bi} = kT \ln [ \frac{ n_n p_p }{n_{i}^{2}} ] \sim 3.25$ V.  The mobile carrier densities $n_n$ and $p_p$  far from the junction in the neutral regions of the diode should be used instead of the doping densities $N_D$ and $N_A$ for evaluating the built-in voltage correctly.  The peak electric field reached at the junction is $F_{max} \sim 2.2$ MV/cm, which is high by conventional standards.  But because of the combination of a large bandgap, and a long depletion depth ($\sim 40$ nm here), the field is insufficient for appreciable interband tunneling current to flow.  Indeed, this is the reason GaN power diodes and transistors can block very high voltages for power electronics applications.  

Several recent experimental \textcolor{red}{\cite{grundmannPSSC07, simonPRL09, rajanAPL10, takeuchiJJAP13}} and theoretical works \textcolor{red}{\cite{schubertPRB10, djPSSA11, tsaiJLT13, fayJxCDC15}} have discussed that introducing a thin polarization dipole can significantly alter the physics of the p-n junction.  Because III-nitride semiconductor heterostructures boast large spontaneous and piezoelectric polarization, one can tunably control the interband tunneling current by suitable design.  Figures \ref{fig1}(b) and (c) show how introducing thin In(Ga)N polarization dipole layers into a GaN p-n junction alters the energy band diagrams, electric fields, and mobile charge distributions in a fundamental manner.  

Introducing a thin In(Ga)N quantum-well layer of thickness $t_{\pi}$ introduces a polarization dipole of sheet density $\sigma_{\pi}$ at the heterojunctions, resulting in a (constant) electric field $\mathcal{F}_{\pi} = q \sigma_{\pi}/\epsilon_s$ confined in the thin layer, and a potential drop $V_{\pi} = \mathcal{F}_{\pi} t_{\pi}$.  
As shown in Fig \ref{fig1}(b), this electric field points in the {\em same direction} as the built-in junction field of the p-n junction in a N-polar crystal with the p-layer on the right.  Since the introduction of an additional layer cannot change the net built-in potential, the area under the electric field curve $\int_{-\infty}^{+\infty} F(x) dx = V_{bi}$ must be unchanged.  This means the depletion region in the p- and n-GaN regions must shrink.  For example, in Fig \ref{fig1}(b), a 2 nm In$_{0.4}$Ga$_{0.6}$N layer causes a peak electric field to reach $\sim 9$ MV/cm, which shrinks the depletion region and makes the GaN band edges appear to have a quasi type-II or staggered band-offset.  Then there must exist the critical condition  
\begin{equation}
t_{\pi} \sigma_{\pi} \geq \frac{\epsilon_s kT}{q^2} \ln [ \frac{ n_n p_p }{n_{i}^{2}} ],
\end{equation}
for which $V_{\pi} \geq V_{bi}$, i.e., the polarization-induced voltage drop {\em exceeds} the built-in junction potential.  Then, instead of a depletion region, the junction is forced to become an {\em accumulation} region.  This is indicated in Fig \ref{fig1}(c), which has a 2 nm InN polarization dipole.  The formation of accumulation regions of electrons and holes [Fig \ref{fig1}(c)] means that the GaN band edges across the junction line up as a Type-III or broken-gap offset.  The maximum fields in the polarization dipole are in the $\sim 20$ MV/cm range, and the field has changed {\em direction} in the adjacent GaN regions.   
\begin{figure}[htbp]
	\centering
		\includegraphics[scale=0.28]{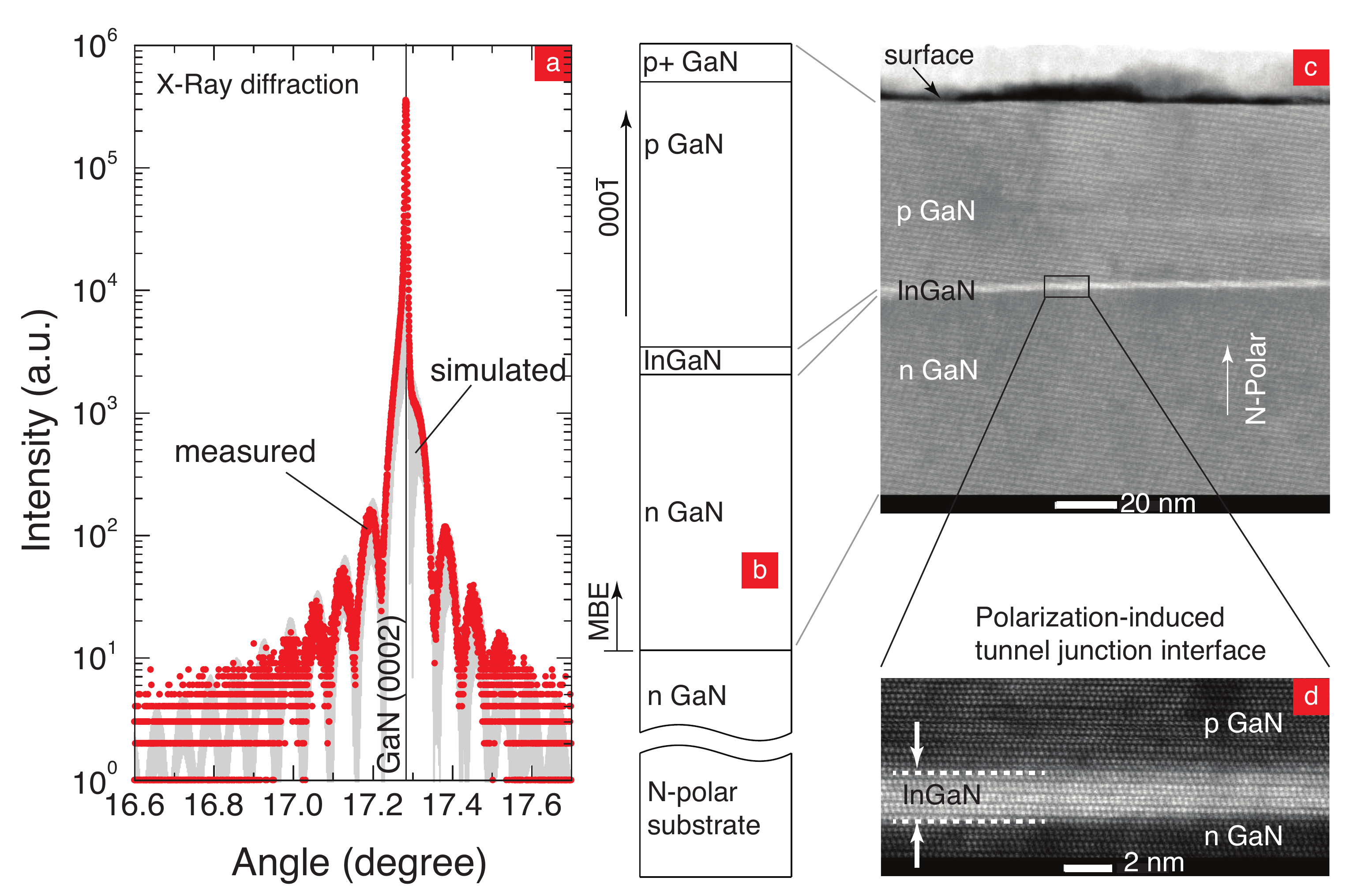}
	\caption{(a) Interference fringes measured (red dots) and simulated (gray) for a MBE-grown p-GaN/InGaN/n-GaN tunnel heterojunction.  The fringes allow estimation of the Indium content.  (b) A schematic cross-section of the heterostructure, (c) Cross-section TEM image, and (d) high-resolution image showing the successful incorporation of a $t_\pi \sim 2$ nm thick InGaN polarization dipole.}
	\label{fig2}
\end{figure}
Such polarization-induced III-nitride heterojunctions make a traditionally wide-bandgap material system rather counter-intuitively attractive for tunneling field-effect transistors \textcolor{red}{\cite{fayJxCDC15}}.  Because of high In composition InGaN, the bandgap (tunneling barrier) is significantly reduced, the depletion region is non-existent, and the electric field is boosted because of polarization.  Polarization charges do not suffer from discrete dopant variations because of their quantum origins in the Berry-phase \textcolor{red}{\cite{vanderbiltPRB93, restaRMP94, bernardiniPRB97, niuRMP10}}.  The fortunate convergence of these effects can be exploited if thin, strained InGaN layer of high Indium composition can be grown in GaN p-n junctions.  The rest of this work presents an experimental study of the effect of increasing Indium composition on interband tunneling currents, and of temperature dependence and degradation of these polarization-induced tunnel junctions.

{\bf {\em Experiment}}: A series of p-GaN/InGaN/n-GaN heterostructures with varying Indium compositions were grown by plasma-assisted molecular beam epitaxy (MBE).  A $\sim$300 nm heavily Si-doped n-GaN layer was grown on N-polar n-type GaN substrates with a nominal dislocation density $\sim 5 \times 10^{7}$/cm$^2$ [Fig \ref{fig2}(b)] at $\sim 700 $ $^{\circ}$C under metal rich conditions. This was followed by a growth interrupt for lowering the substrate temperature and consuming excess Gallium, and then a $t_{\pi} \sim 2$ nm thick InGaN polarization dipole layer was grown at a much lower substrate temperature of $\sim 550$ $^{\circ}$C.  A $\sim$60 nm Mg-doped p-GaN layer with Mg doping density $N_A \sim 2\times 10^{18}$/cm$^{3}$ was grown after the InGaN layer.  The growth of this p-layer started at 550 $^{\circ}$C without an interrupt, and the temperature was gradually ramped up after the InGaN layer was buried.  This prevented the InGaN layer from decomposing, and allowed Mg incorporation with reduced defect density \textcolor{red}{\cite{simonPSSA08}}.  This transition from the InGaN layer to a high-quality p-type doped GaN layer remains a challenge in the MBE growth, because raising the temperature for the optimal p-GaN layer causes the decomposition of the InGaN layer.  The surface was capped with a 5 nm thick heavily Mg-doped p-cap layer for contacts.  The In compositions were varied over four samples with $x_{In} = 0 \rightarrow 0.15 \rightarrow 0.22 \rightarrow 0.25$, with the first serving as the control p-n junction. Fig. \ref{fig2}(a) shows the X-ray diffraction pattern with cavity resonance fringes due to the existence of an ultra thin InGaN layer for the $x_{In}=0.25$ sample; no such fringes are seen in the control sample (not shown).  Fig. \ref{fig2}(b) shows the layer structure, and Figs \ref{fig2}(c, and d) show the transmission electron microscope (TEM) images of the pGaN/InGaN/nGaN heterojunction.  The thicknesses and uniformity of the layers are verified from the image.  The InGaN layers are crystalline, and some defect formation was seen in the p-type GaN layers for the high Indium composition junctions. 

\begin{figure}[]
	\centering
		\includegraphics[scale=0.162]{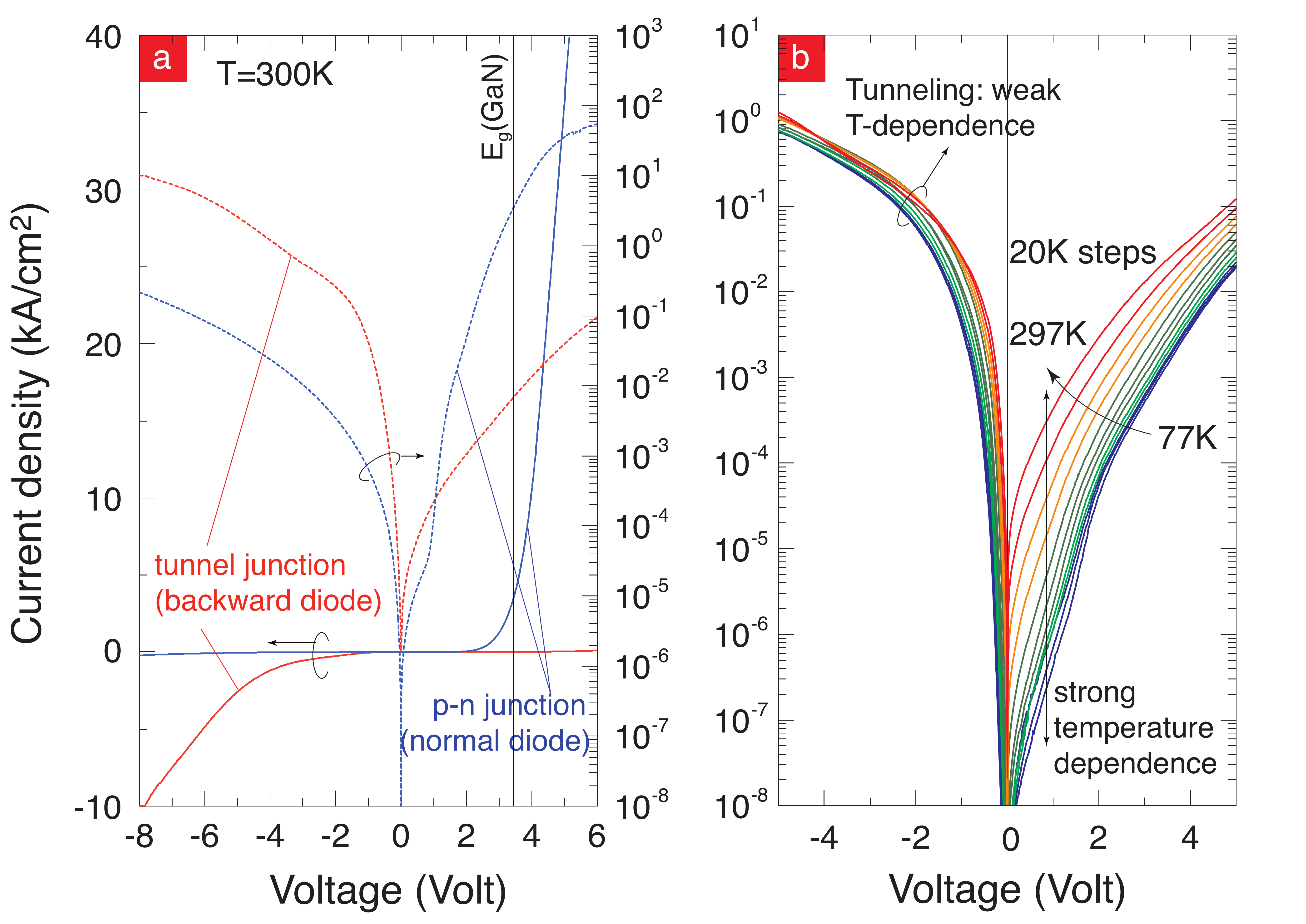}
	\caption{(a) Measured $I-V$ characteristics of a control p-n junction (blue) and the $x_{In}=0.25$ polarization-induced tunnel-junction (red) in linear (solid) and log (dashed) scales.  The tunnel-junction shows backward-diode behavior as a signature of interband tunneling.  (b) Temperature dependence of $I-V$ curves for the p-GaN/InGaN/n-GaN tunnel junction with $x_{In}=0.25$.  The weak temperature dependence of reverse-bias current supports interband tunneling as the transport mechanism. }
	\label{fig3}
\end{figure}

The samples were then processed into pGaN/InGaN/nGaN tunnel diodes by reactive ion etching (RIE) down to the n-GaN  layer, and depositing Ti/Au and Ni/Au metal stacks for n- and p-contacts.  The measured current-voltage (I-V) characteristics at 300 K for the $x_{In}=0.25$ tunnel diode and the control p-n diode are shown in Fig. \ref{fig3}(a).  The control p-n diode is rectifying and turns-on at a voltage close to the bandgap of GaN, as expected.  It also shows strong near band-edge electroluminescence at forward bias (not shown).  The tunnel junction on the other hand shows a strong backward-diode behavior, conducting much higher current at reverse bias than forward bias.  This is an indication of interband tunneling in such junctions \textcolor{red}{\cite{simonPRL09, rajanAPL10}}.  The current densities reach 10s of kA/cm$^2$ at high reverse biases.  

\begin{figure}[]
	\centering
		\includegraphics[scale=0.165]{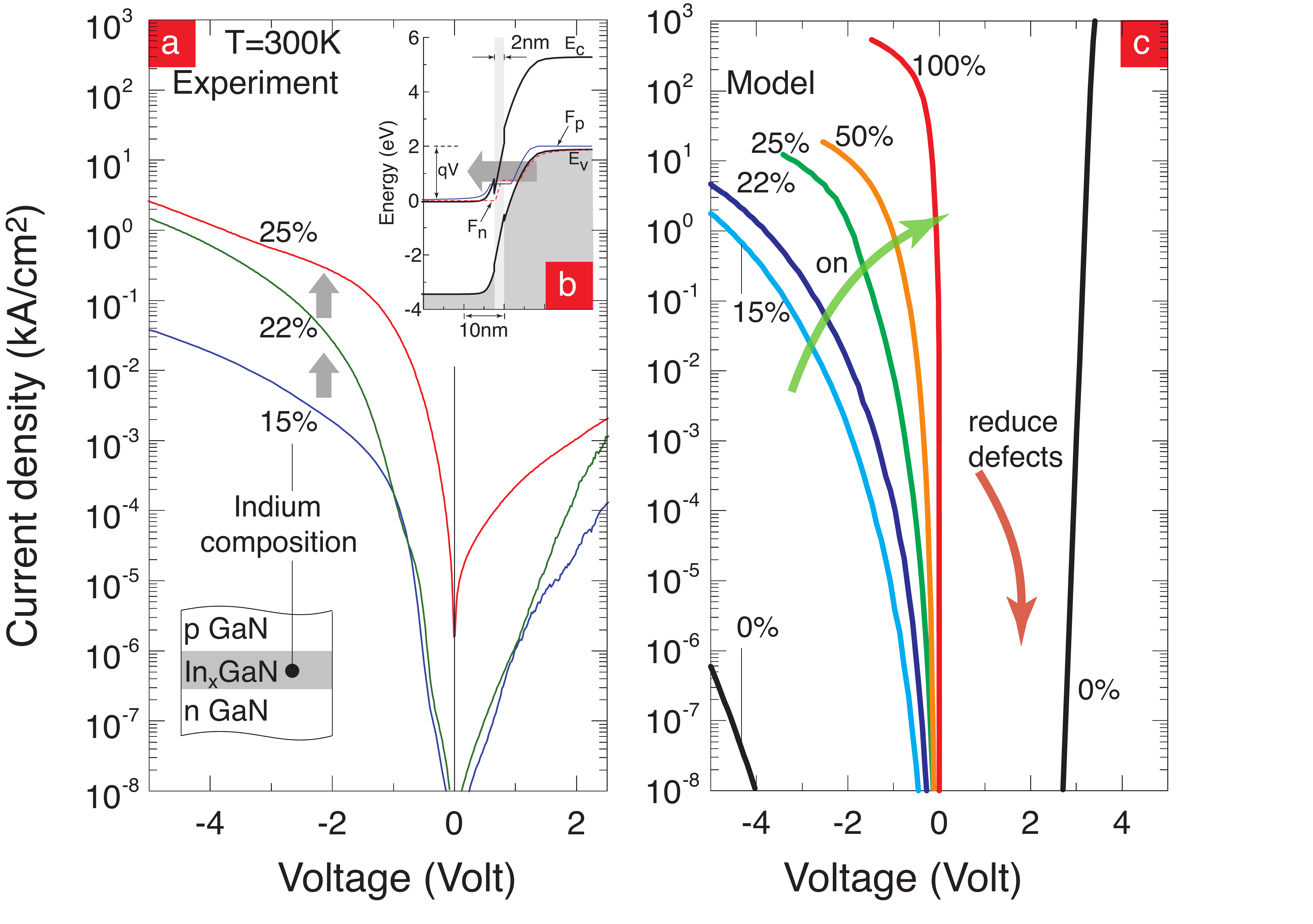}
	\caption{(a) The effect of increasing polarization charge and smaller bandgaps by increasing the Indium composition $x_{In}$ in the tunnel heterojunction.  The reverse bias tunneling current increases with $x_{In}$.  (b) Calculated non-equilibrium energy-band diagram at a bias of $-2$ V, showing the tunneling window of energies. (c) Calculated reverse-bias currents for various In$_x$Ga$_{1-x}$N polarization dipoles, with $0\% \leq x \leq 100\%$.  The forward bias current for the control p-n diode is shown, and remain of similar orders of magnitude for the Indium compositions for which the calculations converged.}
	\label{fig4}
\end{figure}

The tunneling current component is expected to be weakly dependent on temperature to the extent of thermal smearing of the Fermi occupation function.  Fig. \ref{fig3}(b) shows the measured temperature-dependent $I-V$ curves of the tunnel junction with $x_{In}=0.25$ measured from $T= 77 \rightarrow 297$ K, with 20 K increments.  The reverse bias current density has a weak temperature dependence, which is indicative of interband tunneling as the transport mechanism.  In comparison, the forward bias current density has a much stronger temperature dependence.  The ideal forward bias current in a p-n junction is due to the thermionic emission of carriers over the barrier, and exhibits a characteristic exponential temperature dependence.  However, the forward bias currents measured in Fig \ref{fig4}(a) are also significantly influenced by defects and trap states.  In our structures, these defects are predominantly expected to be in the p-GaN layer because of the low-temperature growth conditions.

The effect of the Indium composition on the reverse-bias tunneling current was examined, as shown in Fig \ref{fig4}(a).  Based on the earlier discussion, the interband tunneling current is expected to increase with increasing $x_{In}$, which is clearly observed.  All three tunnel-junctions operate as backward diodes, conducting more in reverse bias than forward bias.  The insert Fig \ref{fig4}(b) shows the numerically calculated non-equilibrium energy band diagram for the $x_{In}=0.2$ tunnel junction at a reverse bias voltage of 2 Volts.  The arrows in Fig \ref{fig4}(a) indicate the increase of current density with In composition.    

Fig \ref{fig4}(c) shows the {\em calculated} current density for various polarization induced tunnel junctions with compositions ranging from $0 \% \rightarrow 15\% \rightarrow 22\% \rightarrow 25\% \rightarrow 50\% \rightarrow 100\%$ Indium in the In$_x$Ga$_{1-x}$N layer.  The tunnel diodes have been simulated with a commercial drift-diffusion based TCAD package (Synopsys Sentaurus)\cite{sentaurus}. In the simulation, the interband tunneling is treated by a WKB (Wentzel-Kramers-Brillouin)-based nonlocal band-to-band generation/recombination model.  Both the reverse-bias tunneling current and the normal forward bias currents are calculated; the forward bias characteristics of the control p-n homojunction is shown.   As can be seen, the increase in the tunneling current with the Indium composition $x_{In}$ is in qualitative agreement with the numerical simulations of current densities.  It is clear that significantly higher $x_{In}$ than obtained here is necessary to achieve $\sim 10^6$ A/cm$^2$ currents at small voltages.  

The reduction of defects by improved growth conditions, and the incorporation of increasing Indium compositions in the InGaN layer are the major challenges moving forward.  This may require going from planar junctions to nanowire or other geometries that allow the incorporation of significant strain due to high $x_{In}$.  But the benefits of achieving such structures go beyond the TFETs and associated devices.  Successful realization of such heterostructures will help probe the electronic bandstructure, polarization fields, and band offsets.  These heterostructures can potentially host exotic electronic states: for example, the strong mixing of the electron and hole states across the InN layer in Fig \ref{fig1}(c), combined with the very high electric field-induced Rashba effect causes an effective band-inversion - as highlighted by the broken gap alignment.  In these structures, topological edge-states are predicted to occur \textcolor{red}{\cite{cvdwPRL12}}.  The search for such exotic electronic states of matter offer strong motivation to address the significant challenges in realizing these heterostructures in the future.  This work was supported in part by the Center for Low Energy Systems Technology (LEAST), one of the six SRC STARnet Centers, sponsored by MARCO and DARPA.

\end{document}